\documentclass[Afour,sageh,times]{arXiv}

\usepackage{moreverb,url}

\usepackage[colorlinks,bookmarksopen,bookmarksnumbered,citecolor=red,urlcolor=red]{hyperref}
\usepackage{graphics} 
\usepackage{epsfig} 
\usepackage{amsmath} 
\usepackage{amssymb}  
\usepackage{color}
\usepackage[table,xcdraw,dvipsnames]{xcolor}
\usepackage{dirtree}
\usepackage{enumitem}
\usepackage{natbib}
\usepackage[normalem]{ulem}
\usepackage{ifthen}

\newcommand\BibTeX{{\rmfamily B\kern-.05em \textsc{i\kern-.025em b}\kern-.08em
T\kern-.1667em\lower.7ex\hbox{E}\kern-.125emX}}

\renewcommand{\sout}{\bgroup \ULdepth=-.45ex \ULset} 
\newcommand\suppress[1]{} 
\newcommand\myOptElse[3]{\ifthenelse{\boolean{#1}}%
{#2\suppress{#3}}%
{\suppress{#2}#3}}

\newboolean{showchanges} 
\setboolean{showchanges}{false}

\newenvironment{removedenvironment}{\color{SkyBlue}}{}
\newcommand*\removed[1]{\myOptElse{showchanges}{\begin{removedenvironment}\sout{#1}\end{removedenvironment}}{}}

\newcommand*\addedapply[1]{{\color{ForestGreen}\textbf{#1}}}

\newcommand*\added[1]{\myOptElse{showchanges}{\addedapply{#1}}{#1}}

\newcommand*\alfasite{https://doi.org/10.1184/R1/12707963}

\begin{document}

\def\publishedurl{\centering \textbf{{\color{red}{The final published version can be accessed from}} \url{https://doi.org/10.1177/0278364920966642}}}

\runninghead{Keipour, Mousaei and Scherer}

\title{ALFA: A Dataset for UAV Fault and Anomaly Detection}

\author{Azarakhsh Keipour, Mohammadreza Mousaei and Sebastian Scherer}

\affiliation{Robotics Institute, Carnegie Mellon University, USA}

\corrauth{Azarakhsh Keipour, AIR Lab,
Robotics Institute,
Carnegie Mellon University,
Pittsburgh, PA 15213,
USA}

\email{keipour@cmu.edu}

\begin{abstract}
We present a dataset of several fault types in control surfaces of a fixed-wing Unmanned Aerial Vehicle (UAV) for use in Fault Detection and Isolation (FDI) and Anomaly Detection (AD) research. Currently, the dataset includes processed data for 47 autonomous flights with \removed{scenarios for eight different types of control surface (actuator and engine) faults}\added{23 sudden full engine failure scenarios and 24 scenarios for seven other types of sudden control surface (actuator) faults}, with a total of 66 minutes of flight in normal conditions and 13 minutes of post-fault flight time. It additionally includes many hours of raw data of fully-autonomous, autopilot-assisted and manual flights with tens of fault scenarios. The ground truth of the time and type of faults is provided in each scenario to enable evaluation of the methods using the dataset. We have also provided the helper tools in several programming languages to load and work with the data and to help the evaluation of a detection method using the dataset. A set of metrics is proposed to help to compare different methods using the dataset. Most of the current fault detection methods are evaluated in simulation and as far as we know, this dataset is the only one providing the real flight data with faults in such capacity. We hope it will help advance the state-of-the-art in Anomaly Detection or FDI research for Autonomous Aerial Vehicles and mobile robots to enhance the safety of autonomous and remote flight operations further. The dataset and the provided tools can be accessed from \url{\alfasite}.
\end{abstract}

\keywords{Dataset, Fault Detection and Isolation, Anomaly Detection, Unmanned Aerial Vehicles, Autonomous Aerial Vehicles, autonomous robots, mobile robotics, field robotics, flight safety, evaluation metrics, ALFA}

\maketitle

\section{Introduction}

The recent growth in the use of Autonomous Aerial Vehicles (AAVs) has increased concerns about the safety of the autonomous vehicles, the people, and the properties around the flight path and onboard the vehicle. To address the concerns, much research is being done on new regulations, more robust systems are designed, and new systems and algorithms are introduced to detect the potential hardware and software issues. 

Many methods have been introduced to detect hardware issues. These methods can be categorized in several ways: they can be learning-based or not, online or offline, identifying the fault type or detecting the anomaly. Each category has its pros and cons. For example, learning-based methods learn models for different fault types and can predict the faults with high precision. However, they have difficulty detecting new issues and are generally dependent on the availability of a large amount of training data, which is not always the case. \cite{Khalastchi:2018:FDD:3177787.3146389} provide a useful review and comparison of different fault detection methods in robotics.

Collecting flight data from real aircraft to test a new Fault Detection and Isolation (FDI) or Anomaly Detection (AD) method is a difficult task; the hardware is expensive, the tests are time-consuming and imposing some of the fault types can lead to the loss of control of the vehicle. As a result, most of the proposed methods are only tested in simulation (\cite{neural-observer, 1-s2.0-S0967066100000460-main, Khalastchi:2013}). The results reported by these methods may be very different from the real data, making a comparison between these methods with the other methods tested on real flight tests difficult. Even many of the methods tested on the real flight data only report a minimal number of tests (\cite{sensors-17-02243, mahalanobis, sensor-anomaly}) and only a few proposed methods have done a reasonable number of tests on the real flight data (\cite{azarakhsh-icra19, VENKATARAMAN2019365}). Providing a large dataset to the FDI and AD community working on UAVs will open the opportunity to test the proposed methods on real data and to compare the results with other methods.

In this paper, we present the Air Lab Fault and Anomaly (ALFA) Dataset, which currently includes processed data for 47 autonomous flights with scenarios for eight different types of \added{sudden }control surface faults, including engine, rudder, aileron(s) and elevator faults\added{, with 23 of the scenarios focusing on full engine failures}. The processed data consists of a total of 66 minutes of normal flight and 13 minutes of post-fault flight time. The dataset also includes several hours of raw autonomous, autopilot-assisted, and manual flight data with tens of different fault scenarios. The processed data provides the ground truth for the exact time and type of fault in each scenario to help with the evaluation of the new methods. A small portion of this dataset has been used by \cite{azarakhsh-icra19} in the evaluation of a real-time anomaly detection method. The current paper describes the dataset in details and opens the access to the complete set of the processed and raw sequences along with the telemetry and dataflash log data of all the flights. Additionally, we provide a set of helper codes for working with the processed data and helping with the evaluation of the new methods in C++, Python and MATLAB languages. The dataset and the tools can be accessed from \removed{\url{\alfasite}}\added{\cite{keipour_mousaei_scherer_2020}}.

\added{The provided dataset can be useful for many types of FDI and AD methods that do not depend on the prior knowledge of the precise dynamic model of the robot. While the precise dynamics model of the robot is not provided, depending on the requirements of a specific method, some form of model can be estimated from the data or from the general airplane dynamics combined with the hardware description of the robot used for the tests.}

In the next section, the platform used for the collection of the dataset and the changes needed to enable the creation of faults are described; Section~\ref{sec:dataset} explains the details about the data and the usage of the dataset; Section~\ref{sec:metrics} proposes the metrics for evaluation of new methods; finally, Section~\ref{sec:summary} summarizes the paper and proposes ideas for the future work.

\section{The Platform} \label{sec:platform}

\subsection{Hardware Setup}

\begin{figure}[!t]
\centering
    \includegraphics[width=0.48\textwidth]{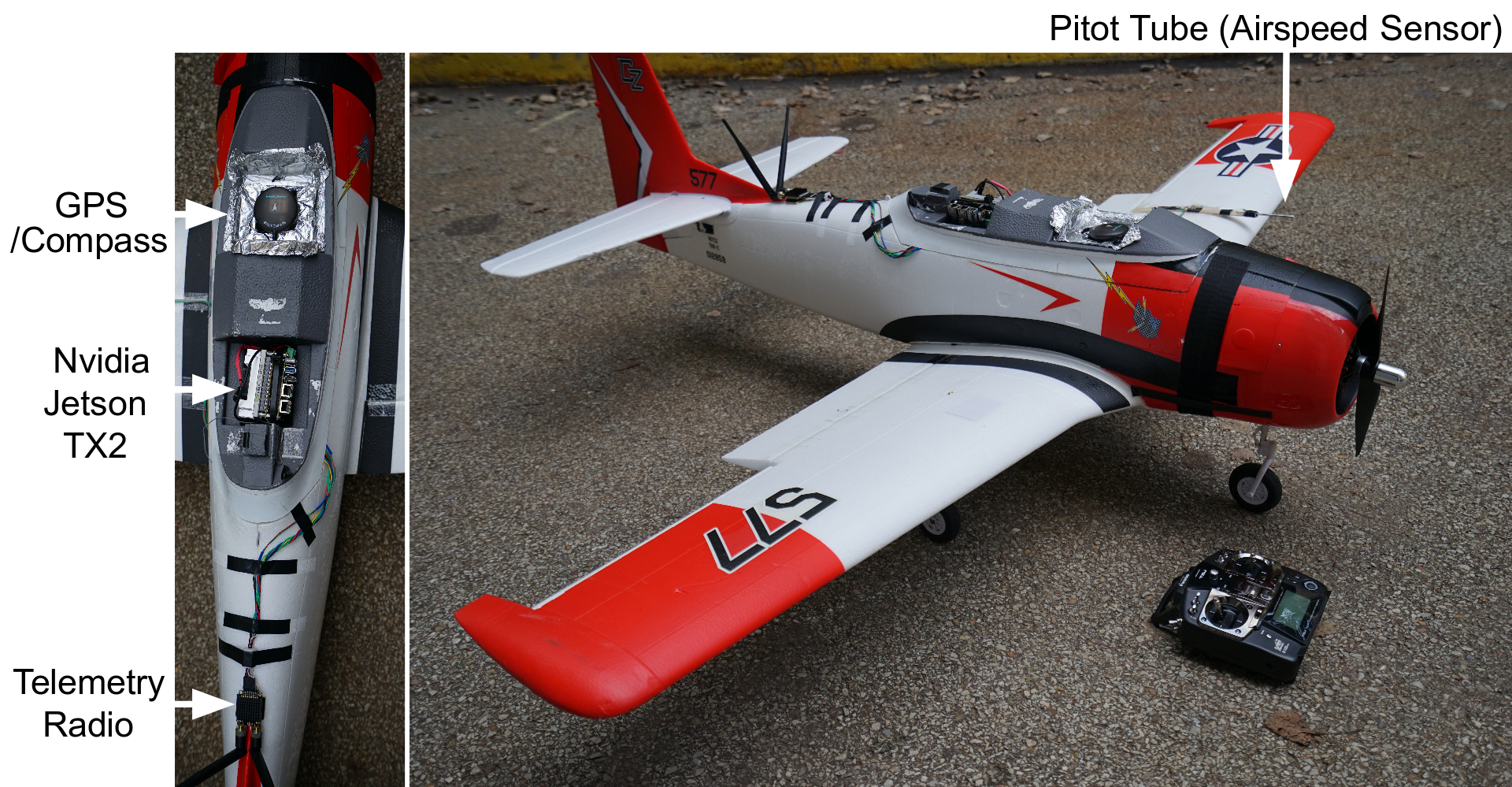}
    \caption{The Carbon-Z T-28 fixed-wing UAV platform equipped with an onboard computer and additional modules for our dataset collection. This is the same platform used by \cite{azarakhsh-icra19} in a previous work.}    
    \label{fig:platform}
\end{figure}

The platform used for the collection of the dataset is a custom modification of the Carbon Z T-28 model plane. The plane has 2 meters of wingspan, a single electric engine in the front, ailerons, flaperons, an elevator, and a rudder. We equipped the plane with a Holybro PX4 2.4.6 autopilot, a Pitot Tube, a GPS module, and an Nvidia Jetson TX2 onboard computer. In addition to the receiver, we also equipped it with a radio for communication with the ground station. 
Figure~\ref{fig:platform} shows the described platform. 

\subsection{Software} \label{sec:software}

The Pixhawk autopilot uses a custom version of Ardupilot/ArduPlane firmware to control the plane in both manual and autonomous modes and to create the simulations. The original firmware is modified from ArduPlane v3.9.0beta1 to include four new parameters as follows:

\textit{DisableEngine}: This parameter can disable the engine to simulate a complete engine failure;

\textit{DisableElevator}: This parameter can fix the elevator in the horizontal position to simulate the stuck elevator;

\textit{DisableRudder}: It can fix the rudder all the way to the left, right or in the middle to simulate the rudder hardover;

\textit{DisableAileron}: It can fix the left aileron, right aileron, or both in the horizontal position to simulate the stuck aileron(s).

\begin{figure}[!t]
\centering
    \includegraphics[width=0.48\textwidth]{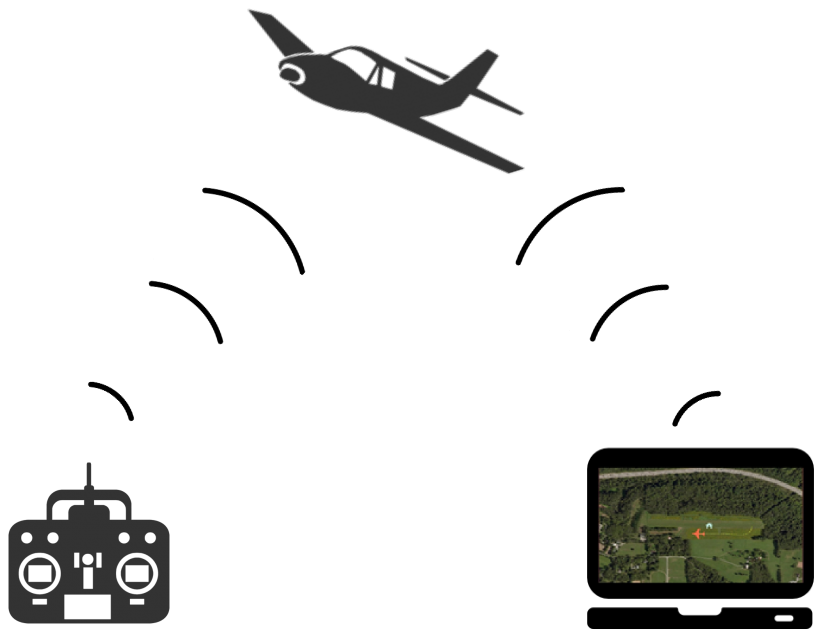}
    \caption{The communication between the safety pilot, the UAV and the ground control station (GCS). The pilot only takes over when the safety is going to be compromised. The GCS is used for disabling the desired control surfaces in the autonomous mode.}
    \label{fig:setup}
\end{figure}

For safety reasons, all the parameters are programmed to only work in the autonomous mode; at any time during the autonomous flight, the safety pilot can take over the control of the plane and all the disabled actuators and the engine will start working normally again. The commands for disabling the control surfaces (modifying the mentioned parameters in the autopilot) can only be sent through the ground control station (GCS). Figure~\ref{fig:setup} shows the communication between the pilot, plane and the GCS.

The onboard computer uses Robot Operating System (ROS) Kinetic Kame on Linux Ubuntu 16.04 (Xenial) to read the flight and state information from the Pixhawk using MAVROS package (the MAVLink node for ROS). The data is recorded as a rosbag, and the ground truth about the faults is periodically published by a node which checks the status of the mentioned custom parameters. The autonomous flight uses a trajectory controller modified from the work of \cite{azarakhsh-icuas18} to enable the control using the onboard TX2 computer instead of the ground station.

Furthermore, to access information about the internal commands of the autopilot (e.g., commanded roll/pitch), both the firmware and MAVROS are modified to publish the desired information in high frequency using the MAVLink protocol. 

\section{Dataset} \label{sec:dataset}
\begin{figure}[!t]
\centering
    \includegraphics[width=0.48\textwidth]{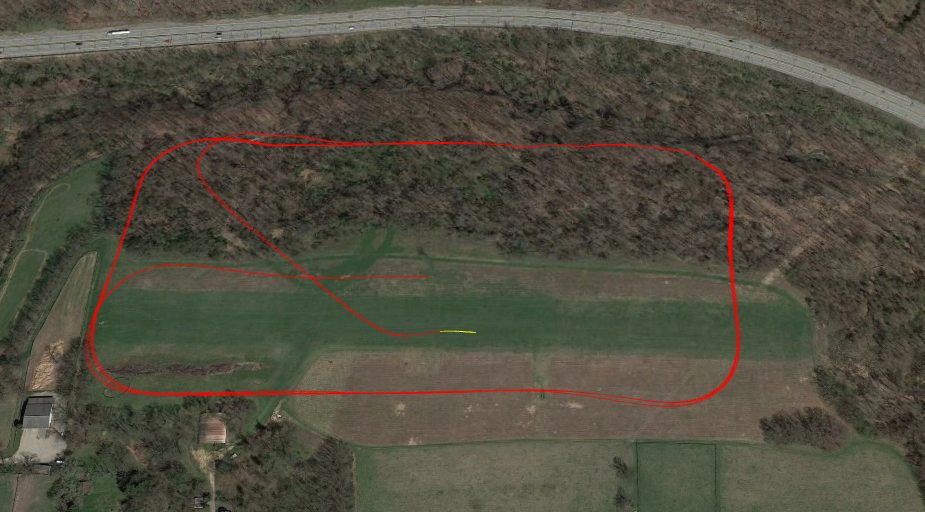}
    \caption{The location of the data recorded and the trajectory used in some of the flight tests.}   
    \label{fig:nardo}
\end{figure}

The presented dataset is entirely collected in an airport around Pittsburgh, Pennsylvania. Figure~\ref{fig:nardo} shows the location of the tests as well as a sample trajectory used in the recorded autonomous flights. Each flight sequence usually includes only a portion of the full trajectory, which can be extracted from the data.

This section describes the data in the dataset in more details, lists the types of faults that are in the dataset, and discusses the provided tools to work with the data and to evaluate an FDI or anomaly detection method using the data.

\subsection{Data Formats}

The dataset consists of 4 types of data, described as below:

\begin{itemize} [leftmargin=*]
\renewcommand\labelitemi{--}
    \item \textit{Autonomous flight sequences with failures}: Flight sequences processed to only contain autonomous flight data and to include failure ground truth data only when there is a fault. Each file contains a flight sequence with at most one fault. The data is provided in three formats: Comma-Separated Values (\texttt{csv}), MATLAB's \texttt{mat}, and the original ROS \texttt{bag}.
    
    \item \textit{Raw flight sequences}: Flight data for flights in all the modes without any preprocessing and is only provided in the original ROS \texttt{bag} format. Some files may include multiple failure test scenarios, while the others may contain no autonomous flight at all. All the files from the first category are cut from these files.
    
    \item \textit{Telemetry logs from TX2}: All the telemetry data is recorded by the onboard TX2 computer from the tests without any preprocessing. The files do not contain the fault ground truth information and can be useful for unsupervised detection methods. More information about the format is available on the ArduPilot website.\footnote{\url{http://ardupilot.org/plane}}
    
    \item \textit{Dataflash logs from Pixhawk}: All the data recorded on the Pixhawk autopilot from the tests without any preprocessing. The files do not contain the fault ground truth information and can be useful for unsupervised detection methods. More information about the format is available on the ArduPilot website.
\end{itemize}

\begin{figure}[!t]
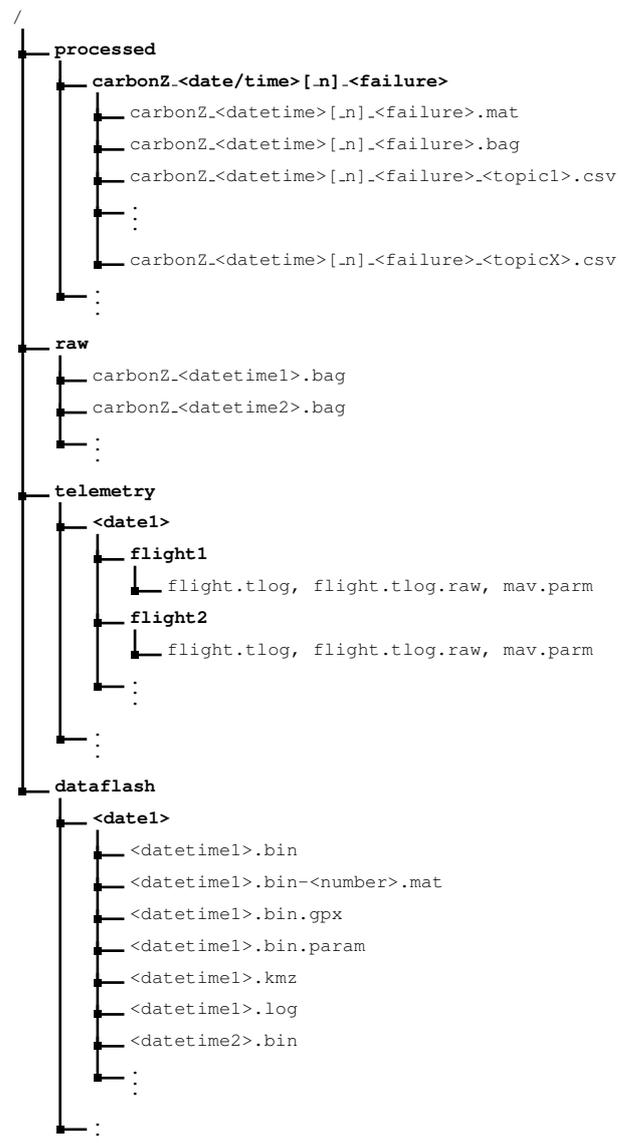

    \DTsetlength{0.2em}{1em}{0.2em}{1pt}{3pt}
    \renewcommand*\DTstyle{\ttfamily \scriptsize}
    \dirtree{%
    .1 /.
    .2 \textbf{processed}.
    .3 \textbf{carbonZ\_<date/time>[\_n]\_<failure>}.
    .4 carbonZ\_<datetime>[\_n]\_<failure>.mat.
    .4 carbonZ\_<datetime>[\_n]\_<failure>.bag.
    .4 carbonZ\_<datetime>[\_n]\_<failure>\_<topic1>.csv.
    .4 \vdots.
    .4 carbonZ\_<datetime>[\_n]\_<failure>\_<topicX>.csv.
    .3 \textbf{\vdots}.
    .2 \textbf{raw}.
    .3 carbonZ\_<datetime1>.bag.
    .3 carbonZ\_<datetime2>.bag.
    .3 \vdots.
    .2 \textbf{telemetry}.
    .3 \textbf{<date1>}.
    .4 \textbf{flight1}.
    .5 flight.tlog, 
    flight.tlog.raw,
    mav.parm.
    .4 \textbf{flight2}.
    .5 flight.tlog, 
    flight.tlog.raw,
    mav.parm.
    .4 \textbf{\vdots}.
    .3 \textbf{\vdots}.
    .2 \textbf{dataflash}.
    .3 \textbf{<date1>}.
    .4 <datetime1>.bin.
    .4 <datetime1>.bin-<number>.mat.
    .4 <datetime1>.bin.gpx.
    .4 <datetime1>.bin.param.
    .4 <datetime1>.kmz.
    .4 <datetime1>.log.
    .4 <datetime2>.bin.
    .4 \vdots.
    .3 \vdots.
    }
    \caption{The directory structure of the dataset.}   
    \label{fig:directory}
\end{figure}

The directory structure of the dataset is shown in Figure~\ref{fig:directory}. The main focus of this dataset is the first data type (the processed flight sequences), and the next few subsections will describe these files in more details. 

\subsection{Fault Types}

The types of the faults currently provided by the dataset are listed in Table~\ref{table:failure-types}. 
\begin{table}[!t]
\centering
\caption{Fault types in the processed dataset.}
\label{table:failure-types}
\resizebox{8.5cm}{!}{%
\begin{tabular}{|
>{\columncolor[HTML]{C0C0C0}} l|c|c|c|}
\hline
\cellcolor[HTML]{CBCEFB}\textbf{\begin{tabular}[c]{@{}c@{}}Fault Type\end{tabular}} & \cellcolor[HTML]{CBCEFB}\textbf{\begin{tabular}[c]{@{}c@{}}\# of Test\\Cases\end{tabular}} & \cellcolor[HTML]{CBCEFB}\textbf{\begin{tabular}[c]{@{}c@{}}Flight Time\\Before Fault (s)\end{tabular}} & \cellcolor[HTML]{CBCEFB}\textbf{\begin{tabular}[c]{@{}c@{}}Flight Time\\w/ Fault (s)\end{tabular}}  \\ \hline
Engine Full Power Loss & 23 & 2282 & 362 \\ \hline
Rudder Stuck to Left & 1 & 60 & 9 \\ \hline
Rudder Stuck to Right & 2 & 107 & 32 \\ \hline
Elevator Stuck at Zero & 2 & 181 & 23 \\ \hline
Left Aileron Stuck at Zero & 3 & 228 & 183 \\ \hline 
Right Aileron Stuck at Zero & 4 & 442 & 231 \\ \hline 
Both Ailerons Stuck at Zero & 1 & 66 & 36 \\ \hline 
Rudder \& Aileron at Zero & 1 & 116 & 27 \\ \hline 
No Fault & 10 & 558 & - \\ \hline \hline
\textbf{Total} & 47 & 3935 & 777 \\ \hline
\end{tabular}
}
\end{table}

As can be seen, a large portion of the dataset is on engine failure, which is provided to help with the Machine Learning-based methods. However, we tried to provide various faults in order to encourage the methods that work on multiple faults. \added{Note that the provided failures still allow the recovery of the robot by the safety pilot and many failure types with a potential of complete loss are not included in the dataset (e.g., the elevator getting stuck all the way down).}

\subsection{Data Description}

The processed file sequences in \texttt{mat} and \texttt{bag} formats include all the available topics, while each \texttt{csv} file includes one topic. 

Each sequence includes information received using the modified MAVROS (as explained in Section~\ref{sec:software}), including the GPS information, local and global state, and wind estimation. Most of the topics are inherited from the original non-modified MAVROS module. These topics are usually available at 4 Hz or higher, and their description can be viewed from the MAVROS website\footnote{\label{mavros-footnote}\url{http://wiki.ros.org/mavros}}. 

In addition to the original MAVROS topics, high-frequency data (between 20 Hz and 25 Hz) is provided on the measured (by sensors) and the commanded (by autopilot) roll, pitch, velocity, airspeed, and yaw. The names of these topics start with \texttt{mavros/nav\_info/}\footnote{\label{footnote-alfa-repo}Custom message type provided with the support code (see Section~\ref{sec:using-dataset}).}; for example, for roll the topic name is \texttt{mavros/nav\_info/roll}. 

At last, the ground truth information is provided as topics for each control surface, starting with \texttt{failure\_status/}. The topics are as follows:

\begin{itemize} [leftmargin=*]
\renewcommand\labelitemi{--}
    \item \texttt{failure\_status/engines}: The value becomes \texttt{true} when engine failure happens.

    \item \texttt{failure\_status/aileron}: The value becomes non-zero when an aileron failure happens. The value of \textit{1} means the failure is on the right side; the value of \textit{2} means that the left aileron is failed; the value of \textit{3} means that both ailerons are failed. Failure here is defined as the aileron(s) getting stuck in zero position.

    \item \texttt{failure\_status/rudder}: The value becomes non-zero when a rudder hardover happens. The value of \textit{1} means that the rudder is stuck in zero position; the value of \textit{2} means that it is stuck all the way to the left; the value of \textit{3} means that the rudder is stuck all the way to the right. 

    \item \texttt{failure\_status/elevator}: The value becomes non-zero when an elevator failure happens. The value of \textit{1} means that the elevator is stuck in zero position; the value of \textit{2} means that it is stuck all the way down. 
\end{itemize}

The ground truth topics appear in a sequence only when there is a fault happening on that control surface. These topics are recorded at about 5 Hz rate; therefore, the first failure ground truth message can happen after up to 0.2 seconds after the exact moment of the fault. 

\begin{figure}[!t]
\centering
    \includegraphics[width=0.48\textwidth]{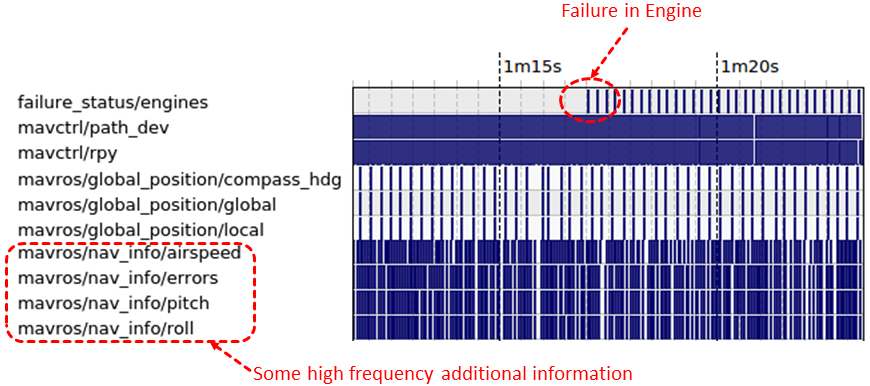}
    \caption{A sample portion of a data file showing the moment that an engine failure happens and some additional high-frequency information provided.}   
    \label{fig:data}
\end{figure}

\added{Table~\ref{table:dataset-fields} provides the list of the topics in the dataset sequences that provide potentially useful information for FDI and AD methods. }Figure~\ref{fig:data} shows a sample of the data for the moment when an engine failure happens. It also shows some of the topics in the data, including the additional data provided by the modified MAVROS.

\begin{table}[!t]
\centering
\caption{{{List of the topics providing potentially useful information in the dataset sequences.}}}
\label{table:dataset-fields}
\resizebox{8.5cm}{!}{%
\begin{tabular}{|
>{\columncolor[HTML]{C0C0C0}} l|c|c|c|}
\hline
\cellcolor[HTML]{CBCEFB}\textbf{\begin{tabular}[c]{@{}c@{}}Field Name\end{tabular}} & \cellcolor[HTML]{CBCEFB}\textbf{\begin{tabular}[c]{@{}c@{}}Description\end{tabular}} & \cellcolor[HTML]{CBCEFB}\textbf{\begin{tabular}[c]{@{}c@{}}Field Type\end{tabular}} & \cellcolor[HTML]{CBCEFB}\textbf{\begin{tabular}[c]{@{}c@{}}Avg.\\Rate\end{tabular}}  \\ \hline
failure\_status/engines & Engine failure status & std\_msgs/Bool & 5 Hz \\ \hline
failure\_status/aileron & Aileron failure type & std\_msgs/Int8 & 5 Hz \\ \hline
failure\_status/rudder & Rudder failure type & std\_msgs/Int8 & 5 Hz \\ \hline
failure\_status/elevator & Elevator failure type & std\_msgs/Int8 & 5 Hz \\ \hline
mavros/nav\_info/airspeed & Airspeed information & mavros\_msgs/NavDataPair \textsuperscript{\ref{footnote-alfa-repo}} & 20-25 Hz \\ \hline
mavros/nav\_info/roll & \begin{tabular}[c]{@{}c@{}}Commanded \& \\measured roll\end{tabular} & mavros\_msgs/NavDataPair \textsuperscript{\ref{footnote-alfa-repo}} & 20-25 Hz \\ \hline
mavros/nav\_info/pitch & \begin{tabular}[c]{@{}c@{}}Commanded \& \\measured pitch\end{tabular} & mavros\_msgs/NavDataPair \textsuperscript{\ref{footnote-alfa-repo}} & 20-25 Hz \\ \hline
mavros/nav\_info/yaw & \begin{tabular}[c]{@{}c@{}}Commanded \& \\measured yaw\end{tabular} & mavros\_msgs/NavDataPair \textsuperscript{\ref{footnote-alfa-repo}} & 20-25 Hz \\ \hline
mavros/nav\_info/errors & \begin{tabular}[c]{@{}c@{}}Tracking, airspeed\\and altitude errors\end{tabular} & mavros\_msgs/NavError \textsuperscript{\ref{footnote-alfa-repo}} & 20-25 Hz \\ \hline
mavros/nav\_info/velocity & \begin{tabular}[c]{@{}c@{}}Commanded \& \\measured velocity\end{tabular} & mavros\_msgs/NavVector3 \textsuperscript{\ref{footnote-alfa-repo}} & 20-25 Hz \\ \hline
mavctrl/path\_dev & \begin{tabular}[c]{@{}c@{}}Path deviation\\(\cite{azarakhsh-icuas18})\end{tabular} & geometry\_msgs/Vector3 & 50 Hz \\ \hline
mavctrl/rpy & \begin{tabular}[c]{@{}c@{}}Measured roll, pitch \\and yaw\end{tabular} & geometry\_msgs/Vector3 & 50 Hz \\ \hline
mavros/global\_position/* & \begin{tabular}[c]{@{}c@{}}Global position info \textsuperscript{\ref{mavros-footnote}}\end{tabular} & various types & 4-5 Hz \\ \hline
mavros/local\_position/* & \begin{tabular}[c]{@{}c@{}}Local position info \textsuperscript{\ref{mavros-footnote}}\end{tabular} & various types & 4 Hz \\ \hline
mavros/imu/* & IMU state \textsuperscript{\ref{mavros-footnote}} & various types & 10 Hz \\ \hline
mavros/setpoint\_raw/* & Setpoint messages \textsuperscript{\ref{mavros-footnote}} & various types & 20-50 Hz \\ \hline
mavros/state & FCU state \textsuperscript{\ref{mavros-footnote}} & mavros\_msgs/State & 1 Hz \\ \hline
mavros/wind\_estimation & \begin{tabular}[c]{@{}c@{}}Wind estimation by FCU \textsuperscript{\ref{mavros-footnote}}\end{tabular} & geometry\_msgs/TwistStamped & 2-3 Hz \\ \hline
mavros/vfr\_hud & \begin{tabular}[c]{@{}c@{}}Data for HUD \textsuperscript{\ref{mavros-footnote}}\end{tabular} & mavros\_msgs/VFR\_HUD & 2-3 Hz \\ \hline
mavros/time\_reference & \begin{tabular}[c]{@{}c@{}}Time reference from \\SYSTEM\_TIME \textsuperscript{\ref{mavros-footnote}}\end{tabular} & sensor\_msgs/TimeReference & 2 Hz \\ \hline
\end{tabular}
}
\end{table}

\subsection{Using the Dataset} \label{sec:using-dataset}

The \texttt{bag} files can easily be played back using the shell commands provided by \texttt{rosbag} ROS package. \added{The custom message definition files are provided to allow working with the topics defined by the custom message types in the \texttt{bag} files.}
Besides, the base tools are provided that allow working with the dataset using C++'11, MATLAB and Python 3.x programming languages. The functionalities include loading a dataset file in memory, iterating through the whole dataset or a single topic in timestamp order, plotting a specific topic field, and some other methods \added{such as separating}\removed{like filtering} the data for normal flight\removed{s} from the post-fault flight. There is no dependency on ROS or any other external package\removed{s}, and all the code is written using the standard libraries of these programming languages.

\added{In addition, a}\removed{The} base code is provided in C++'11 language to help with the evaluation of new fault and anomaly detection methods. It automatically subscribes to the ground truth topics and waits for the method to publish the detection. It then returns information about the false positives, the delay in the detection, and some other statistics.

\section{Evaluation Metrics} \label{sec:metrics}

The metrics used for evaluation of a method can vary based on the method's class, the fault types, and the method applications. For example, for online methods, the delay between the fault happening and the detection can be critical for the safety of the flight, while for the offline methods this metric may not be as important and in many cases (e.g., for fault classification methods) it can be meaningless. We propose the following metrics for the evaluation of methods using the provided dataset:

\begin{itemize} [leftmargin=*]
\renewcommand\labelitemi{--}
    \item \textit{Maximum Detection Time:} For online methods, the delay between the time a fault happens and the time it is detected is an important factor when comparing two methods. In real applications, a large detection delay in any scenario can lead to irreversible situations, resulting in the complete loss of control of the vehicle. Therefore, the maximum detection time is a useful metric for the evaluation of online methods.

    \item \textit{Average Detection Time:} The average detection time over the set of fault scenarios shows the overall time performance of a method in detecting faults and is also a useful metric for the evaluation of the online methods.

    \item \textit{Accuracy:} This metric is the ratio of the number of correctly classified sequences to the total number of sequences. Any false result (false fault detection or not detecting a fault) is considered as a misclassified case. This metric considers all the positive and negative sequences and is suitable to get an overall idea about the performance of an algorithm, but works best when the false detections and false negatives have similar costs.
    
    \item \textit{Precision:} This metric is the ratio of the sequences with correctly predicted faults to the total number of detections (both true and false positive detections). Each sequence containing a false detection counts as a false positive and each sequence containing only correct detection(s) counts as a true positive. This metric indicates how reliable is the method when it announces a fault.

    \item \textit{Recall:} This metric is the ratio of the sequences with correctly predicted faults to the total number of sequences containing fault(s). Each sequence containing only correct detection(s) counts as a true positive. This metric indicates how reliable is the method in detecting the faults.
\end{itemize}


\section{Summary and Future Work} \label{sec:summary}

In this paper, we presented ALFA dataset that provides autonomous flight data of a fixed-wing UAV with scenarios of different control surface faults in the middle of the flights. We believe that this dataset will be highly useful in Fault Detection and Isolation (FDI) and Anomaly Detection (AD) research.

The presented ALFA dataset is the most extensive dataset for fault and anomaly detection in Autonomous Aerial Vehicles, but it is by no means a complete dataset. The dataset contains several sudden control surface failures, but many more types of faults can happen in UAVs, including issues in sensors and gradual errors.  
We invite other groups and researchers to contribute to the dataset by providing their test data with other types of faults and platforms. It will significantly increase the speed of research in this area and will help with benchmarking future methods.

We provided base codes to help with using the dataset and evaluation of the new methods with it. To further extend the usefulness of the data, it will be beneficial to create a benchmark to provide researchers with a better tool to compare their methods with the state-of-the-art.

\section*{Funding}
This work was supported through NASA Grant Number NNX17CL06C.

\begin{acks}
This project became possible due to the support of Near Earth Autonomy (NEA). Also, the authors would like to thank Mark DeLouis for his help as the pilot during the months of testing and recording the data.
\end{acks}

\bibliographystyle{Formatting-IJRR/SageH}

\end{document}